\documentclass[prb,twocolumn, floatfix]{revtex4}

\usepackage{graphicx}
\usepackage{epstopdf}

\begin{document}

\title{Millikelvin thermal and electrical performance of lossy transmission line filters}
\author{D. H. Slichter}
\email[Corresponding author: ]{slichter@berkeley.edu}

\author{O. Naaman}

\author{I. Siddiqi}
\affiliation{Materials Sciences Division, Lawrence Berkeley National Laboratory, Berkeley CA 94720\\Quantum Nanoelectronics Laboratory, Department of Physics, University of California, Berkeley CA 94720}

\date{\today}

\begin{abstract}

We report on the scattering parameters and Johnson noise emission of low-pass stripline filters employing a magnetically loaded silicone dielectric down to 25 mK.  The transmission characteristic of a device with $f_\textrm{\scriptsize -3dB}$=1.3 GHz remains essentially unchanged upon cooling.  Another device with $f_\textrm{\scriptsize -3dB}$=0.4 GHz, measured in its stopband, exhibits a steady state noise power emission consistent with a temperature difference of a few mK relative to a well-anchored cryogenic microwave attenuator at temperatures down to 25 mK, thus presenting a matched thermal load. 

\end{abstract}


\maketitle

Electrical devices probing quantum coherence or operating near the quantum noise limit are very sensitive to the effective temperature and impedance of adjacent circuitry over a broad spectrum of frequencies. Reduced sensitivity and decoherence can result from spurious radio and microwave frequency signals or broadband radiation from components not in thermal equilibrium with the electrical device. Filtering unwanted electromagnetic radiation both in and out of the measurement frequency band is thus imperative in quantum measurement. Conventional reactive filters typically are not suited to this task\textemdash their stopband does not extend over many decades in frequency,\cite{denny} resulting in transmitted noise and presenting an uncontrolled impedance that does not behave as a matched, thermalized load.

For dc measurements, low-pass filters consisting of wires embedded in metal-powder-impregnated resin\cite{martinis,bladh} are routinely used in the sub-kelvin temperature range where many quantum electrical circuits operate. However, their cutoff frequency does not extend to the microwave regime and their impedance is not well-controlled. Several types of cryogenic low-pass filters suitable for radio frequency signals have been developed, including impedance-controlled metal powder filters,\cite{milliken} lossy coaxial cables,\cite{zorin} and lossy stripline filters.\cite{santa}  Two important outstanding questions are first, whether the composite lossy dielectric material, non-magnetic or magnetic, used in such filters retains its frequency-dependent loss at cryogenic temperatures, and second, whether it actually reaches thermal equilibrium with its surroundings. In this letter, we characterize the operation of planar transmission line filters consisting of a copper stripline surrounded by a commercial magnetically loaded silicone dielectric down to millikelvin temperatures. These filters are absorptive in their stopband.  We observe no significant change in their transmission characteristics upon cooling, and that the filters thermalize to within a few millikelvin of a well-anchored nichrome attenuator used as a noise reference down to 25mK.  

\begin{figure}[htbp]
\includegraphics{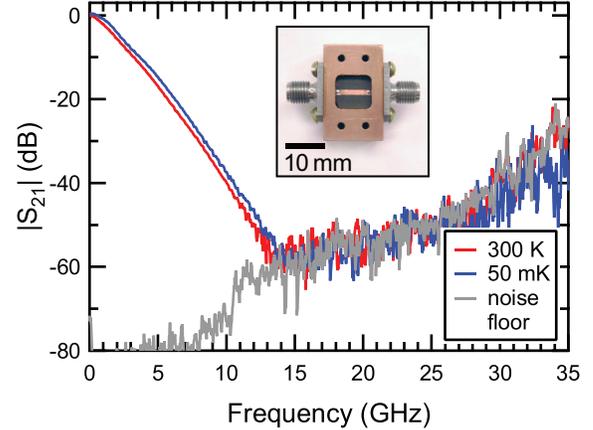}
\caption{\label{fig1} (Color online) Transmission characteristics of filter $\textrm{F}_\textrm{\scriptsize 9.5mm}$ with $f_\textrm{\scriptsize -3dB}$=1.3 GHz at T=300K and T=50 mK.  The dynamic range of the measurement decreases with frequency due to attenuation in the cryogenic coaxial lines. Inset: photo of the filter, with lid and upper dielectric removed.}
\end{figure}

The filters consist of a copper box containing a section of 50 $\Omega$ stripline with Emerson \& Cuming Eccosorb MFS-117 dielectric, as shown in the inset of Fig.\ \ref{fig1}.  The dielectric material has a magnetic loss tangent that increases with frequency.\cite{emerson} Similar types of Eccosorb have been used as black body loads in microwave cosmology experiments and have been shown to thermalize very well at temperatures down to at least 2K.\cite{richards} The width of the center conductor was designed using a series approximation\cite{pozar} for the impedance of a stripline in a cavity with conducting walls, modified to account for a dielectric with nonzero electric and magnetic loss tangents and non-unity relative permeability.\cite{emerson} For a cavity cross-section 8.9 mm wide by 4.2 mm high with a 50 $\mu$m thick, 1.45 mm wide copper center conductor, we calculate an impedance close to 50 $\Omega$ across a broad range of frequencies. This optimal strip width for broadband matching was verified with room temperature S-parameter measurements.    
		
We present data from two filters. The first, $\textrm{F}_\textrm{\scriptsize 9.5mm}$, with a 9.5 mm-long stripline, was designed to have a -3 dB passband of $f_\textrm{\scriptsize -3dB}$=1.3 GHz. The second one, $\textrm{F}_\textrm{\scriptsize 70mm}$, was designed with a 70 mm-long stripline to give significant attenuation at 1.5 GHz and act as a thermal emitter at this frequency.  The filters were anchored to the mixing chamber of a dilution refrigerator which can be operated at temperatures from 25 mK to 500 mK.

We first tested transmission characteristics of our filters with a vector network analyzer, calibrated at room temperature to the reference plane of the connectors on the filter.  Figure \ref{fig1} shows the measured transmission $|S_{21}|$ of $\textrm{F}_\textrm{\scriptsize 9.5mm}$ at room temperature (red) and at 50 mK (blue). The transmission is essentially unchanged at low temperature, and in both cases the transmitted signal is below the analyzer noise floor (grey) at frequencies above 14 GHz.  As the measurement frequency increases, so does the attenuation of the stainless steel and copper-nickel microwave coaxial lines inside the cryostat, resulting in a decrease of dynamic range and an increase of the noise floor. The frequency response agrees well with S-parameters calculated from the filter geometry and loss tangent data from Emerson \& Cuming. The small shift between the 300 K and 50 mK traces is consistent with a slight reduction in the line attenuation upon cooling. Though the frequency dependence of the magnetic loss is weak and results in the slow filter rolloff shown in Figure \ref {fig1}, the advantage of this filter is that it remains lossy up to the highest measured frequency. Similar types of Eccosorb maintain their loss beyond 2 THz.\cite{hemmati}  The passband at low frequencies can be sharpened by integrating a reactive filter in series with the lossy filter.  The measured $|S_{21}|$ is reproducible under repeated thermal cycling, and after complete disassembly and reassembly of the filter.  Filter $\textrm{F}_\textrm{\scriptsize 70mm}$ was designed to be completely absorptive in the 1-2 GHz band and exhibited $|S_{21}| = -25$ dB at 1.5 GHz. 

\begin{figure}
\includegraphics{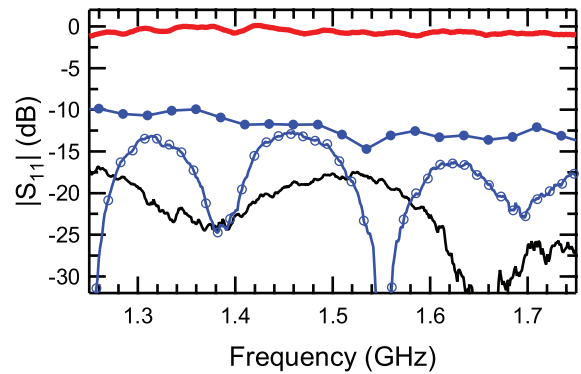} 
\caption{\label{fig2} (Color online) Reflection characteristics at 25 mK from an open circuit (bold red), a 50 $\Omega$ termination (black), $\textrm{F}_\textrm{\scriptsize 70mm}$ with opposite end open-circuited (empty blue circles), and $\textrm{F}_\textrm{\scriptsize 9.5mm}$ (at 50 mK) connected to a 50 $\Omega$ load (filled blue circles).}
\end{figure}

Filter return loss was measured with a standard reflectometry setup with four circulators, two at base temperature and two preceding the 4K HEMT amplifier. Figure \ref{fig2} shows the reflection $|S_{11}|$ from 1.25 GHz to 1.75 GHz of filter $\textrm{F}_\textrm{\scriptsize 9.5mm}$(filled circles), held at 50 mK and terminated with a 50 $\Omega$ load, as well as the longer filter $\textrm{F}_\textrm{\scriptsize 70mm}$(open circles), held at 25 mK and terminated with a reflective short.  We find that both filters are rather well matched to 50 $\Omega$, with $\textrm{F}_\textrm{\scriptsize 9.5mm}$ and $\textrm{F}_\textrm{\scriptsize 70mm}$ exhibiting $|S_{11}|$ less than -10 dB and -13 dB, respectively. Note that this entire frequency range is well within the stopband of $\textrm{F}_\textrm{\scriptsize 70mm}$ and that since the filter is terminated with a short, its return loss is solely due to absorption in the dielectric. To test the accuracy of our calibration, we used a cryogenic coaxial switch to disconnect the filters under test and connect an open circuit (red) and a 50 $\Omega$ load (black), both at 25 mK. We observe near unit reflection from the open and a good broadband match from the load, as expected.

These results indicate that the attenuation of Eccosorb MFS-117 is unchanged in magnitude and frequency dependence when cooled to millikelvin temperatures. We now demonstrate that this material does indeed thermalize at these temperatures by measuring the noise power emitted from $\textrm{F}_\textrm{\scriptsize 70mm}$ well in its stopband. A direct measurement of the Johnson noise emitted from the filter is hampered by $1/f$ drifts in the gain and system noise temperature of the measurement chain, which limit the maximum integration time. Given our system noise temperature of 7.4K, these drifts give a maximum measurement resolution of approximately 150 mK.  

\begin{figure}[b]
\includegraphics{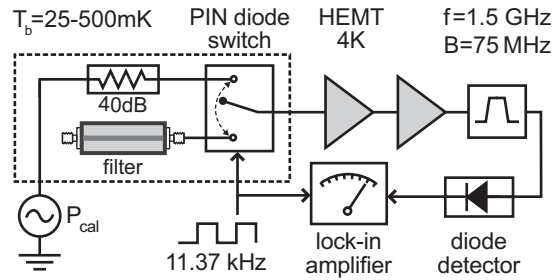}
\caption{\label{fig3} Schematic of the measurement apparatus.  Filter $\textrm{F}_\textrm{\scriptsize 70mm}$ is connected to a PIN diode switch with the other port of the filter shorted.  A 40 dB attenuator is connected to the other input of the switch and to a signal generator at room temperature with output power $\mathrm{P_{cal}}$ via a well-attenuated line (10 dB at 1 K, 20 dB at 4 K, not shown in figure).  The output of the switch passes through four isolators (not shown), amplifiers, and a bandpass filter with center frequency $f=1.5$ GHz and bandwidth $B=75$ MHz.  A zero-bias Schottky diode detector is used to rectify the RF power signal.  The switch is chopped between the attenuator and $\textrm{F}_\textrm{\scriptsize 70mm}$ with a 50\% duty cycle at 11.37 kHz and the diode output voltage is synchronously detected with a lock-in amplifier.}
\end{figure}

To circumvent this problem, we used a differential measurement technique, shown schematically in Fig. \ref{fig3}.  We used a GaAs PIN diode switch located at the mixing chamber of our refrigerator to chop between our filter and a 40 dB nichrome attenuator, both anchored to the mixing chamber plate. The noise from both sources was amplified and then detected with a zero-bias Schottky diode. The difference in noise power emission between the filter and the attenuator was measured synchronously with a lock-in amplifier.  A signal generator connected to the attenuator was used for calibrating the noise power.  Nichrome attenuators are routinely used in measurement of quantum devices, and are believed to thermalize down to 30 mK.\cite{vijay, lehnert} This measurement therefore compares the noise radiated from the filter to that of a well-anchored metallic load.

\begin{figure}[tb]
\includegraphics{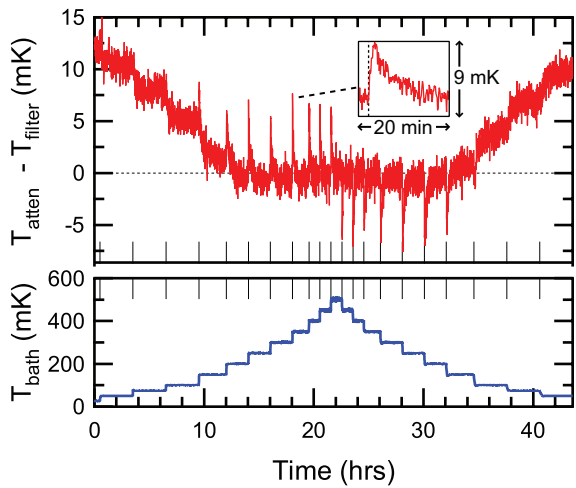}
\caption{\label{fig4}(Color online) Top: temperature difference between attenuator and filter $\textrm{F}_\textrm{\scriptsize 70mm}$ as a function of time.  Bottom: temperature of the mixing chamber plate plotted against the same time axis.  The long tick marks on the horizontal axis demarcate the times when the refrigerator temperature was changed. Top inset: expanded view of a $\Delta T$ spike after a temperature change.  The time of the temperature change is shown with a dotted line.}
\end{figure}

To quantify the temperature difference between the filter and the attenuator, we set $P_\textrm{\scriptsize{cal}}=-135$ dBm and measured the in-phase quadrature of the lock-in signal $V_\textrm{\scriptsize{lockin}}=\eta Gk_BB(T_\textrm{\scriptsize{atten}}-T_\textrm{\scriptsize{filter}})$, where $\eta$ is the power to voltage rectification ratio of the diode, $G$ is the gain of the measurement chain, $B=75$ MHz is the bandwidth, $T_\textrm{\scriptsize{atten}}$ and $T_\textrm{\scriptsize{filter}}$ are the noise temperatures of the attenuator and filter respectively, and $k_B$ is Boltzmann's constant.  To extract the temperature difference, we used two independent calibration procedures. First, after each data point acquisition, five calibration points with different values of $P_\textrm{\scriptsize{cal}}$ ranging from -63 dBm to -78 dBm were recorded. In this case, the lock-in signal is $V_\textrm{\scriptsize{lockin}}=\eta G[AP_\textrm{\scriptsize{ca}l}+k_BB(T_\textrm{\scriptsize{atten}}-T_\textrm{\scriptsize{filter}})]$ where $A=-76$ dB is the attenuation of the all stainless steel coaxial input line (measured independently at room temperature).  From the $V_\textrm{\scriptsize{lockin}}=0$ intercept of this linear relation we find $\Delta T\equiv T_\textrm{\scriptsize{atten}}-T_\textrm{\scriptsize{filter}}$, independent of the system gain. As a crosscheck, we connected the PIN diode switch to the 50 $\Omega$ attenuator, varied its temperature from 200 mK to 600 mK, and recorded the diode output voltage. This yielded the product $\eta G$ which can also be used to obtain $\Delta T$.  These two methods yield values of $\Delta T$ that agree to within 30 \%. Since the first calibration procedure is carried out for every data point and the latter only once for the entire measurement, this close agreement indicates a high degree of immunity to $1/f$ noise in the measurement. Additionally, we measured the insertion loss of both channels of the PIN diode switch and calculate the effects of transmission imbalance and noise emission to be negligible. 
	
In Figure \ref{fig4}, we plot the measured $\Delta T$ as a function of time, together with the temperature of the thermal bath as it is varied between 25 mK and 500 mK. Our data indicate that the attenuator and the filter have temperatures within a few millikelvin of each other down to 25 mK.  At high temperatures ($>200$ mK), the measured temperature difference between the attenuator and the filter is within the experimental uncertainty, and each temperature step is accompanied by a spike in $\Delta T$: a sharp increase followed by a slower decrease, shown in the inset of Fig. \ref{fig4}.  We interpret this as the attenuator rapidly thermalizing to the new bath temperature, followed by the filter equilibrating with a longer time constant. Note that the spike polarity is reversed when we step the temperature downwards. Below 150 mK, the filter appears to be slightly colder than the attenuator, even after several hours of measurement.  

	 In summary, we have demonstrated that lossy transmission line filters made with Eccosorb MFS-117 magnetically loaded dielectric do indeed retain their frequency dependent loss and have comparable thermal response to a nichrome attenuator down to 25 mK. These filters are effective at suppressing high frequency electromagnetic noise without radiating non-equilibrium thermal noise, present a well-matched 50 $\Omega$ impedance, and are thus well suited for measurements involving quantum devices. 

DHS gratefully acknowledges support from a Hertz Foundation Fellowship endowed by Big George Ventures. This work was supported by the Laboratory Directed Research and Development Program of Lawrence Berkeley National Laboratory under Department of Energy Contract No. DE-AC02-05CH11231.

\bibliography{thermal.bib}

\end{document}